\documentclass[aps,pra,twocolumn,superscriptaddress]{revtex4-1}

\usepackage{amsmath}

\usepackage{bm}
\usepackage{amssymb,amsfonts,latexsym,fancyhdr,graphicx,epstopdf,times,txfonts}

\usepackage{graphicx}

\usepackage{overpic}
\usepackage[english]{babel}
\usepackage{graphicx}
\usepackage {subfigure}
\usepackage[pdftex,colorlinks=true,allcolors=blue]{hyperref}

\usepackage{xcolor}

\usepackage{verbatim}

\begin{document}


\title{Spectral properties of correlated quantum wires and carbon nanotubes \\ within the Generalized Kadanoff-Baym Ansatz}

\author{F. Cosco}
\affiliation{Institut f\"ur Theoretische Physik und IQST, Albert-Einstein-Allee 11, Universit\"at Ulm, D-89081 Ulm, Germany}
\author{N. W. Talarico}
\affiliation{QTF Centre of Excellence, Turku Centre for Quantum Physics, Department of Physics and Astronomy, University of Turku, 20014 Turku, Finland}
\author{R. Tuovinen}
\affiliation{QTF Centre of Excellence, Turku Centre for Quantum Physics, Department of Physics and Astronomy, University of Turku, 20014 Turku, Finland}
\author{N. Lo Gullo}
\affiliation{QTF Centre of Excellence, Turku Centre for Quantum Physics, Department of Physics and Astronomy, University of Turku, 20014 Turku, Finland}

\selectlanguage{english}

\begin{abstract}
We investigate the spectral properties of an open interacting system by solving the Generalized Kadanoff-Baym Ansatz (GKBA) master equation for the single-particle density matrix, namely the time-diagonal lesser Green's function. 
To benchmark its validity, we compare the solution obtained within the GKBA with the solution of the Dyson equation (equivalently the full Kadanoff-Baym equations).  In both approaches, we treat the interaction within the self-consistent second-order Born approximation, whereas the GKBA still retains the retarded propagator calculated at the Hartree-Fock level. 
We consider the case of two leads connected through a central correlated region where particles can interact and exploit the stationary particle current at the boundary of the junction as a probe of the spectral features of the system. 
In this work, as an example, we take the central region to be a one-dimensional quantum wire and a two-dimensional carbon nanotube and show that the solution of the GKBA master equation well captures their spectral features. Our result demonstrates that, even when the propagator used is at the Hartree-Fock level, the GBKA solution retains the main spectral features of the self-energy used.
\end{abstract}

\maketitle
\section {Introduction}
In recent years there has been a growing interest into the properties of correlated systems under external perturbations; the latter being continuous drivings~\cite{Foieri2010,Sentef2015,Purkayastha2017,Kalthoff2019,Honeychurch2019}, coupling with macroscopic reservoirs with whom they can exchange 
energy and particles~\cite{Myohanen2012, Latini2014, Antipov2017, Ridley2018, Ridley2019, Covito2020, Cohen2020, Dutta2020}, or strong external pulses with a finite duration in time~\cite{Freericks2009,Eckstein2013,Sentef2013,Schueler2016,Mor2017,Perfetto2019}. The field of application of these studies is broad, encompassing out-of-equilibrium phases~\cite{Murakami2017,Sentef2017,Li2018,Sentef2018,Topp2018}, pump-probe experiments and time-resolved dynamical properties~\cite{Fausti2011,Denny2015,Mitrano2016,Werdehausen2018}, band-gap and Floquet engineering~\cite{Lindner2011,Wang2013,Mahmood2016,Huebener2017,Kennes2019,Topp2019}, transport in correlated systems~\cite{Kaiser2014,Hu2014,McIver2020,Talarico2020}, equilibration and thermalization in strongly correlated materials~\cite{Eckstein2011,Ligges2018,Peronaci2018} and quantum gases~\cite{LoGullo2016,Settino2020TG}, and relaxation in nano-structures~\cite{Cassette2015,Kemper2018}.
Despite such a wide range of applications, the theoretical description of out-of-equilibrium many-body systems remains a challenging task. This difficulty arises because different, and relevant, ingredients need to be included in the description. Some of these are many-body interactions, external time-dependent fields and the possibility of exchanging energy and matter with the environment. All of these elements are essential to obtain a reliable description of the observed phenomena and/or to give solid ground to new predictions.

In this spirit, several new system-specific approaches have been proposed, and more established ones have been modified and improved. Each of them, however, comes with its own advantages as well as limitations.  Some of the most popular are the time-dependent density-matrix-renormalization group (TD-DMRG) and the related tensor-network methods \cite {Schollwock2011}. They have the appealing feature of treating many-body interactions in an essentially exact way but the inclusion of coupling to external leads is often realized with the inclusion of effective baths \cite{Nusseler2020}. However, complications arise when considering initial system-environment correlations and high-dimensional geometries. Another, completely different, class of numerical methods is based on perturbative  approaches suited to account for external time-dependent fields or many-body interactions. Among these, we recall the non-equilibrium Green's functions (NEGFs), the dynamical mean field theory (DMFT), and the time-dependent density functional theory (TD-DFT) \cite{Stefanucci2004}. All of them are perturbative in some parameter, either the many-body interaction, the coupling to the leads, or the tunnelling energy within the system. These techniques allow to include  system-environment correlations in a somehow straightforward way. Moreover their application is virtually unaffected by the system geometry and dimensionality for their computational complexity does not scale with the size of the Hilbert space but rather with the space dimension

\begin{figure}[t!]
	\includegraphics[width=\linewidth]{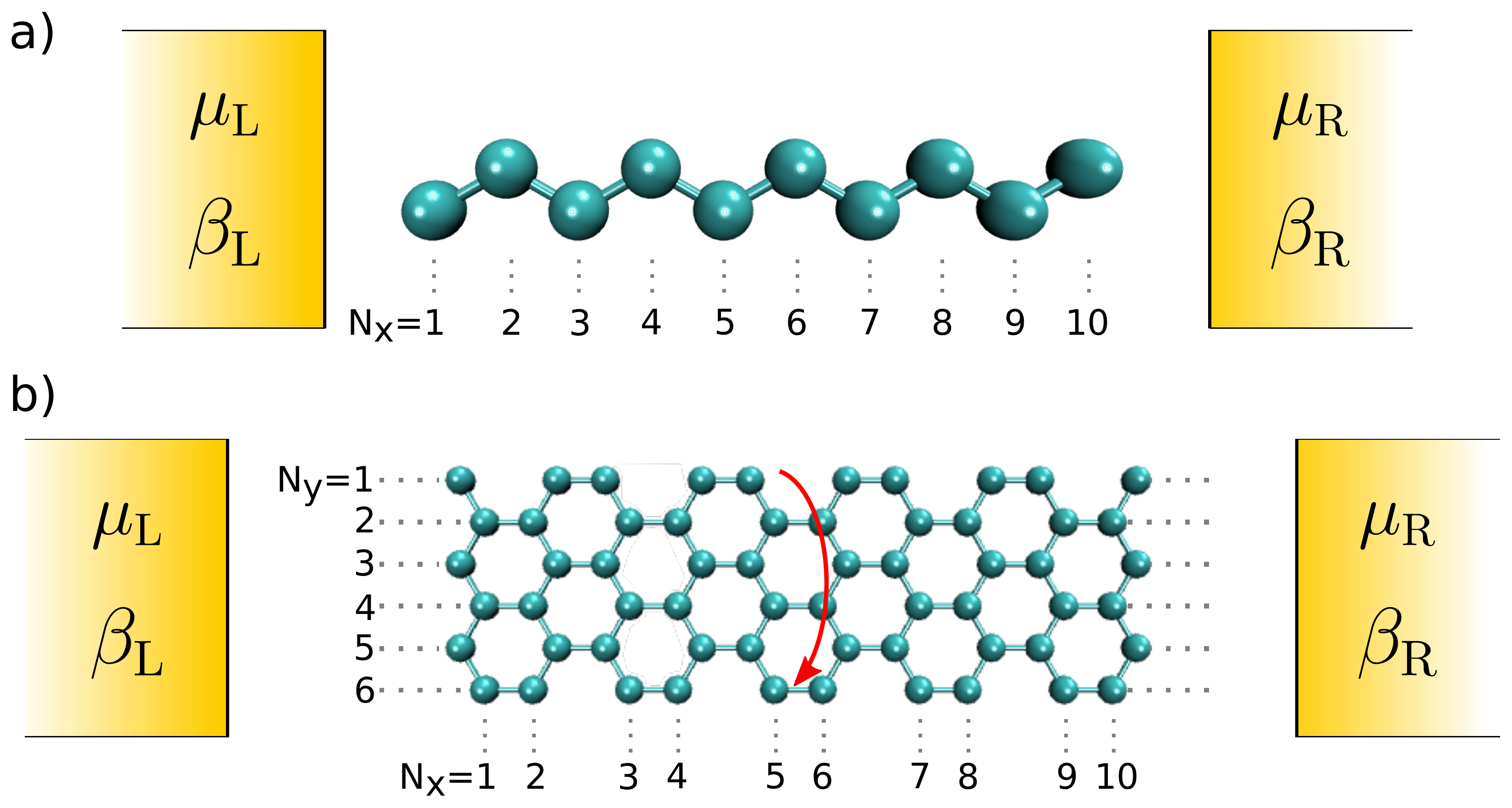}
	\caption{(Color online) Schematic of the transport setups considered. {\bf a)} A quantum wire connected with two metallic leads. Only the outer-most sites are connected to the leads. {\bf b)} A carbon nanotube connected to two metallic leads. The left-most carbon atoms are connected to the left lead, and the right-most carbon atoms are connected to the right lead (rows $N_y=\{1,3,5\}$). The red arrow signifies periodic boundary condition along the $y$-direction folding a graphene nanoribbon into a carbon nanotube.}
	\label{fig:CNT}
\end{figure}

In particular, the non-equilibrium Green's function approach, although in principle very well suited to study the dynamics of open interacting many-body systems, is computationally very demanding. This is due to the two-time structure of the Kadanoff-Baym equations (KBEs), or alternatively of the Dyson equation. The computational cost of the NEGF can be lowered by resorting to the so called Generalized Kadanoff-Baym Ansatz (GKBA)~\cite{Lipavsky1986, Spicka2005, Balzer2013gkba, Hermanns2012, Hermanns2013}, which lifts the two-time structure of the equations and allows to derive a master equation for the system's density matrix.
The GKBA was proposed to derive an equation of transport for quantum systems which would include quantum correlations, allowing to go beyond the Boltzmann equation. Similarly to the Keldysh approach, it was overlooked for a long time due to the complexity of the equations to be solved.  One of the limitations of this approach is the fact that the spectrum of the system is often computed at the Hartree-Fock (mean-field) level. This might induce to think that the solution of the GKBA cannot capture spectral features beyond the Hartree-Fock approximation even when the master equation includes higher-order corrections as the second-order Born approximation.
Attempts to go beyond this approach have been proposed~\cite{Haug1996,Bonitz1996,Bonitz1999,Kwong1998,Pal2009,Latini2014,Hopjan2018,Perfetto2018CHEERS}, which nevertheless may compromise the numerical advantage of the GKBA over the KBEs. Interesting and promising progresses have been made both in the inclusion of initial correlations~\cite{Karlsson2018,Hopjan2019,Bonitz2019} and in the possibility of widening the allowed many-body perturbation schemes~\cite{Schlunzen2017,Schlunzen2019}. 
Furthermore, a second reformulation that maps the GKBA integro-differential equations onto a coupled system of ordinary differential equations has been proposed and has lead to linear scaling of the scheme with the computational time~\cite{Schlunzen2020, Joost2020, Karlsson2020}.

In this work, we show that the GKBA is able to capture fundamental features of the spectrum of a correlated many-body system when the stationary particle current is used as a probe for such properties. Specifically, we look at two different systems and compare the results of the stationary state of the GKBA-HF master equation with the solution of the stationary state of the full Dyson equation~\cite{Talarico2019}. The latter is solved in the frequency domain and not in the two-time plane because here we are not interested in comparing the transient dynamics. We choose to study two systems with different dimensionality, in particular a one-dimensional system, representing a correlated quantum wire, and a more realistic two-dimensional one, representing a carbon nanotube, see Fig.~\ref{fig:CNT}. With the help of these systems we are able to show that our findings, and therefore the application of the GKBA itself, do not depend upon the dimensionality of the system.

Although we analyze a transport setup, our conclusions that the GKBA is able to retain some features of the self-energy approximation, in our case the second Born, is a general result and is valid in different physical platforms. Specifically we show that these features are encoded into the time-off-diagonals of the lesser Green's function despite the fact that they are computed with the Hartree-Fock propagator. This aspect make the GKBA a valuable numerical tool to be employed in the description of experimental setups where the lesser Green's function is used to compute the measured signal such as the time resolved angle-resolved photo emission spectroscopy (t-ARPES) or the pump-and-probe technique.

\section {GKBA: Open Interacting Systems}
\label{sec:GKBA}
In this section, we give a brief overview of the NEGF approach and how the GKBA is introduced within this framework. Although this procedure is well established, we want to highlight some features of the GKBA master equation which will then lead us to the main point of our work.
For the sake of definiteness, we consider a fermionic system, interacting via a two-body interaction, coupled to a fermionic bath, see Fig.~\ref{fig:CNT}. The dynamics of such a system is described by the following second-quantized time-dependent Hamiltonian
\begin {equation}
\begin{split}
\hat H(t) = \sum_{i,j}^{} h_{ij}(t) \hat c^\dagger_i \hat c_j
+\frac{1}{2}\sum_{ijkl}u_{ijkl}(t) \hat{c}_i^\dagger \hat{c}_j^\dagger \hat{c}_k \hat{c}_l\\
+  \sum_{i,k,\alpha} T_{ik}^{\alpha}(t) \hat c^\dagger_i \hat d_{\alpha,k}+ T_{ik}^{\alpha*}(t) \hat d^\dagger_{{\alpha ,k}} \hat c_i + \sum_{\alpha,k} \epsilon_{\alpha,k}  \hat d^\dagger_{\alpha,k} \hat d_{\alpha,k},
\end{split}
\label {tot-ham}
\end {equation}
where $\hat c^\dagger_i,  \hat c_j$ are the creation (annihilation) operators of the correlated central system, while $\hat d^\dagger_{\alpha, k},  \hat d_{\alpha,k}$ are the creation (annihilation) operators of infinite baths, with $\alpha$ labeling each environment. Hence, $h_{ij}(t)$ is the time-dependent single-particle Hamiltonian, $u_{ijkl}(t)$ the two-body interaction tensor, and $T^{\alpha}_{ik}(t)$ is the time-dependent coupling matrix between the modes of the system and the modes of each environment $\alpha$.

Within the NEGF formalism the primary object is the single particle Green's function (SPGF) defined on the Keldysh contour as 
\begin{equation}
\label{eq:spgf}
G_{ij} (z; z') = -\mathrm {i}  \left\langle \mathcal {T}_{c} \hat c_{i} (z) \hat c_{j}^\dagger (z')\right\rangle
\end{equation}
where $z$ and $z'$ are complex variables and $\mathcal T_c$ is the time-ordering operator which orders time over the Schwinger-Keldysh contour.

The Green's function satisfies the equation of motion
\begin{equation}
(\mathrm i \partial_z-h(z))G_{} (z;z') =\delta_c (z,z)+ \int \mathrm d\bar z \,  \varSigma (z;\bar z) G (\bar z;z'),
\label{eq:KBE-contour}
\end{equation}
where we have introduced $\varSigma(z;z')$ as the self-energy operator which accounts for the two-body interaction and the coupling between system and environment. The self-energy has two contributions $\varSigma = \varSigma_{\mathrm {MB}}+\varSigma_{\mathrm {emb}}$.  The first term is the many-body self-energy, containing the effects of the interaction between particles within the system, and the second term is the embedding self-energy, accounting instead for the coupling of the system to the leads. This latter term describes the exchange of matter and /or energy between the leads and the system. 

By means of the  Langreth rules, we obtain  from Eq. \eqref {eq:KBE-contour} the equations of motion for the lesser and greater real-time components of the single-particle Green's function in Eq.~\ref{eq:spgf} (we neglect the vertical imaginary track):
\begin{align}
(\mathrm i \partial_t-h(t))G^{\lessgtr} (t;t') = I^{\lessgtr}(t,t'), \label{eq:KBE-time1}\\
G^{\lessgtr} (t;t')(-\mathrm i \stackrel{\leftarrow}{\partial}_{t'}-h(t')) = I^{\lessgtr}(t,t'),
\label{eq:KBE-time2}
\end{align}
where the collision integrals are given by:
\begin{align}
 I^{<}( t, t')=\int \mathrm d\bar t \, \varSigma^{<
} (t;\bar t) G^{A} (\bar t;t') + \varSigma^{R} (t;\bar t) G^{<} (\bar t;t')\\
 I^{>}( t;t')=\int \mathrm d\bar t \, G^{<
} (t;\bar t) \varSigma^{A} (\bar t;t') + G^{R} (t;\bar t) \varSigma^{<} (\bar t;t').
\label{collint}
\end{align}
The retarded/advanced component of the Green's function are instead obtained by solving
\begin{equation}
(\pm \mathrm i \partial_t-h(t))G^{R/A} (t;t') = \delta(t,t')+ \int \mathrm d\bar t \, \varSigma^{R/A
} (t;\bar t) G^{R/A} (\bar t;t').
\label {eq:propagator}
\end{equation}
Together, Eqs. \eqref {eq:KBE-time1}, \eqref {eq:KBE-time2} and \eqref {eq:propagator} are part of the Kadanoff-Baym equations (KBEs), which, in the general case also contain the equations of motion for the right and left component of the Keldysh Green's function~\cite{Stefanucci2013}. The solution of the KBEs is computationally demanding, especially for large systems and/or long times due to the double-time structure of the objects involved.

The Generalized Kadanoff Baym Ansatz (GKBA) was introduced to reduce the complexity of the Kadanoff-Baym equations and the computational cost necessary to solve them \cite{Lipavsky1986}. Loosely speaking, the key idea underlying this approach is to decouple the dynamics of the time-diagonal components of the SPGF, namely to the single-particle density matrix of the system $\rho (t)=- \mathrm i G^{<} (t,t)$, from the off-diagonal ones. 
The master equation for the density matrix reads:
\begin{equation}
\frac {d} {dt}  {\rho (t)}  + \mathrm {i} \left [h_{HF}(t),\rho (t) \right ] = - (I (t) +{\rm h.c.}),
\label{eq:master-equation}
\end{equation}
where we have defined the Hartree-Fock (HF) Hamiltonian $h_{HF,ij} (t)= h(t)+ \sum_{m,n} w_{imnj} (t) \rho_{nm} (t)$, with $w_{imnj} (t) = 2 u_{imnj}(t)- v_{imjn} (t)$ to account for the mean-field effects. The collision integral is given by
\begin{equation}
I(t) = \int_0^t \mathrm{d} \bar{t} [ \varSigma^>(t,\bar{t}) G^<(\bar{t},t) - \varSigma^<(t,\bar{t}) G^>(\bar{t},t) ],
\end{equation}
and it contains the exchange self-energy that accounts for the effects of the interactions beyond the mean-field.
The calculation of the collision integral requires the knowledge of the greater and lesser Green's functions $G^\lessgtr (t,t')$ at different times.
In order to reduce the computational cost, the latter quantities are approximated with the first term of the formal solution of the Dyson series, i.e.
\begin{equation}
G^{\lessgtr}(t,t') \approx \mathrm {i} \left [G^{R} (t,t') G^{\lessgtr} (t',t')-G^{\lessgtr} (t,t) G^{A} (t,t')  \right],
\label{eq:lg-GKBA}
\end{equation}
where they now depend exclusively on the retarded/advanced Green's functions and the single-particle density matrix, as $G^{>}(t,t)=-\mathrm i ( {1}- \rho(t))$. For the forthcoming discussion, it is worth to mention that for the approximation in Eq. \eqref {eq:lg-GKBA}, the fundamental identity $G^>-G^<=G^R-G^A$ still holds.
To close the ansatz a suitable approximation for the retarded/advanced propagators has to be provided. The main requirement is that this does not have to be more computationally costly than solving the GKBA master equation for the reduced single-particle density matrix. For closed systems, the most common choice satisfying this requirement is to compute the retarded Green's function at the Hartee-Fock (HF) level. Hereafter we will refer to the resulting approximation as the GKBA-HF. Other possibilities have been studied and put forward~\cite{Bonitz1996,Bonitz1999,Pal2009} which allow to go beyond the HF approximation and which are at the same time more computationally convenient than solving Eq.~\eqref{eq:propagator}. 
The GKBA-HF approach has already been successfully applied to describe closed many-body systems~\cite{Hermanns2014, Schlunzen2017, Bostrom2018, Perfetto2018, Tuovinen2019pssb, Schueler2019, Perfetto2019, Perfetto2020, Murakami2020, Schueler2020}, and more recently its application to the dynamics of open quantum systems has been studied~\cite{Latini2014,Hopjan2018,Bostrom2019}. For open systems a similar problem arises due to the embedding self-energy. Once again, in order to avoid solving Eq.~\eqref{eq:propagator} fully, it is possible to resort to the wide band limit approximation (WBLA) for the embedding self-energy~ \cite{Latini2014}. The latter has the same beneficial effect as the HF approximation for the many-body self-energy, namely it is local in time and thus is a delta function in the two-time plane.
In what follows, we choose the propagator in the HF+WBLA approximation which is the formal solution of Eq. \eqref {eq:propagator} and thus given by
\begin{equation}
\begin{split}
G^{R/A}(t,t') &=\mp \mathrm {i} \theta [\pm( t-t')]  T e^{ - \mathrm i \int_{t}^{t'} \mathrm d \bar t \, (h_{HF} (\bar t) - \mathrm i \Gamma/2 ) }\\ &=\mp \mathrm {i} \theta [\pm( t-t')]   Y(t,t'),
\label{G_ret}
\end {split}
\end{equation}
where we have defined the matrix $\Gamma_{ij}= \sum_{\alpha} \Gamma_{ij}^\alpha$, with $\Gamma_{ij}^\alpha=\int d\omega\sum_k\delta(\omega-\epsilon_k)T_{ik}^{\alpha}(T_{jk}^{\alpha})^*$, and the auxiliary operator $Y(t,t')= T e^{ - \mathrm i \int_{t}^{t'} \mathrm d \bar t \, (h_{HF} (\bar t) - \mathrm i \Gamma/2 ) }$. 
The lesser/greater components of the embedding self-energy are still needed to calculate the collision integral and their expression is given by
\begin{equation}
\begin{split}
\varSigma_{\mathrm {emb},ij}^<  (t,\bar t) &= \mathrm {i} \sum_{\alpha} \Gamma_{ij}^\alpha  \int  \mathrm {d} \omega \, f_\alpha (\beta, \mu,  \omega)e^{-\mathrm i \omega(t-\bar t)},\\
\varSigma_{\mathrm {emb},ij}^>  (t, \bar t) &= \mathrm {i} \sum_{\alpha}\Gamma_{ij}^\alpha  \int \mathrm {d} \omega \, (1-f_\alpha (\beta, \mu, \omega))e^{-\mathrm i \omega(t-\bar t)},
\end {split}
\end{equation}
where $f_\alpha (\beta, \mu, \omega)$ is the Fermi-Dirac distribution of the the environment $\alpha$, and depends on the inverse temperature $\beta$ and chemical potential $\mu$ of each bath. \\
In the GKBA, for the many-body part of the system Hamiltonian, the HF propagator includes the effect of the interaction at the mean-field level in the time-diagonal component, whereas the collision integral is used to account for higher order contributions. In this work, we employ the second order Born approximation (2B), for which the lesser/greater self-energies are~\cite{Hermanns2012,Latini2014,Tuovinen2019-2B,Settino2020}
\begin{equation}
\varSigma_{\mathrm {MB},ij}^{2B,\lessgtr}  (t,\bar t)= \sum_{mnpqrs} u_{irpn} (t)w_{mqsj} (\bar t) G^{\lessgtr}_{nm}(t,\bar t) G^{\lessgtr}_{pq}(t,\bar t) G^{\gtrless}_{sr}(\bar t, t).
\label{se-2b}
\end{equation}
The self-energies in Eq. \eqref {se-2b} are advantageously calculated within the GKBA framework, as they do not require time integrals. By means of the Green's function components defined in  Eq. \eqref {eq:lg-GKBA} and Eq. \eqref {G_ret} the $2B$ self-energies become
\begin{equation} 
\begin{split}
\varSigma_{\mathrm {MB},ij}^{2B,\lessgtr}  (t,\bar t)= \sum_{mnpqrs} u_{irpn} (t)w_{mqsj} (\bar t) [Y(t,\bar t)G^{\lessgtr}(\bar t,\bar t)]_{mn}\\
\times [Y(t,\bar t)G^{\lessgtr}(\bar t,\bar t)]_{pq} [G^{\lessgtr}(\bar t,\bar t) Y(\bar t,t)]_{sr}.
\end{split}
\end{equation}
Under the assumptions made, wide-band-limit for the environmental degrees of freedom, and within the 2B approximation the GKBA retains a computational cost scaling as  $\sim t^2$, with $t$ the simulation time.
The approach is therefore very promising as it allows to explore the long time dynamics of large systems retaining at the same time some of the most appealing features of the NEGFs such as the inclusion of correlations beyond the mean-field approach in the collision integrals.

\section{Spectral properties of the GKBA master equation and particle current}
\label{sec:curr}

Although its appealing features the GKBA approach comes with some limitations which inhibit its use to a wider range of physical systems. The first limitation is the inclusion of correlations in the initial state of the system. This is usually solved through an initial preparation phase in the simulation which decreases the effective useful time. A second major limitation comes from the choice of the self-energies embodying the features of interaction-generated correlations in the system. Related to this latter point there is a third one: inclusion of the correlation effects into the single particle spectrum given by the retarded Green's function.
The limitation of the GKBA master equation in capturing the spectral features of correlated many-body systems seems conceptually more difficult to be overcome.  Any attempt to include correlations in the retarded Green's function would require the solution of Eq.~\eqref{eq:propagator} with a self-energy not local in time, thus frustrating any computational advantage of the GKBA. This unavoidably means that the single-particle spectrum embodied in the spectral function 
\begin{equation}
\mathcal A(\omega)=\underset{T\rightarrow \infty}{\lim}\int d\tau\;  \mathcal A(T+\tau/2,T-\tau/2),
\label{eq:spfunc}
\end{equation}
with the center-of-time coordinate $T=(t+t')/2$ and the relative-time coordinate $\tau=t-t'$. Because $ \mathcal A(t,t')=G^R(t;t')-G^A(t;t')$ it does not contain any correlation. This is why it is often stated~\cite{Hopjan2019} that the solution of the GKBA master equation cannot capture spectral features beyond the Hartree-Fock approximation even when the collision integral includes higher-order effects for the many-body self-energy. 
It seems therefore that the GKBA betrays the promises of lowering the computational cost of the KBE in simulating the spectral and dynamical features of correlated many-body systems while maintaining the same order of approximation.

Nonetheless the spectral properties of a system are reflected into physically relevant quantities. This is why the spectral function is so important in the first place: its knowledge helps to explain physical properties of a system and predict the behavior of physical quantities. 
Hereafter we address the case of a transport setup in which two non-interacting electronic leads are connected through a central conducting region where many-body interactions take place. An applied bias voltage across the junction can make currents develop between the electronic reservoirs. 
This current and the electric conductivity are the physical quantities which we are going to look at and connect them with spectral properties of the central region.
The general expression for the particle current that flows into the lead $\alpha$ is given by:~\cite{Meir1992, Jauho1994, Tuovinen2013}
\begin{equation}
I_\alpha(t)=2\textrm{Re}\left\{\int d\tau\; \textrm{Tr}\left[\varSigma_\alpha^<(t,\tau)G^A(\tau;t)+\varSigma_\alpha^R(t,\tau)G^<(\tau;t)\right]\right\}.
\label{eq:curr-time}
\end{equation}
In long-time limit and in the absence of external drive, the above expression can be rewritten in the frequency domain as:
\begin{equation}
	I_\alpha^{(S)}=i\int \frac{d\omega}{2\pi}\; \textrm{Tr}\left[\varSigma_\alpha^<(\omega)\mathcal A(\omega)-\Gamma_\alpha(\omega) G^<(\omega)\right],
\label{eq:curr}
\end{equation}
where $\Gamma_\alpha(\omega)=i\left(\varSigma_\alpha^R(\omega)- \varSigma_\alpha^A(\omega)\right)$ and $\mathcal A(\omega)=i\left( G^R(\omega)- G^A(\omega)\right)$. This expression shows manifestly the dependence of the current from the spectral function $\mathcal A(\omega)$ and the Fourier transform of the lesser Green's function $ G^<(\omega)$.
Eq.~\eqref{eq:curr} can be further simplified and it is usually rewritten in the Landauer-B\"uttiker form~\cite{Landauer1957, Buttiker1986}
\begin{equation}
I_\alpha^{(S)}=\int \frac{d\omega}{2\pi}\; \sum\limits_{\beta}T_{\alpha \beta}(\omega)(n_\alpha(\omega)-n_\beta(\omega)),
\label{eq:LB}
\end{equation}
with the transmission coefficient defined as $T_{\alpha \beta}(\omega)=\textrm{Tr}\left[\Gamma_\alpha(\omega) G^R(\omega)\Gamma_\beta(\omega) G^A(\omega)\right]$.
Although this form has a more intuitive and immediate physical meaning, it could suggest that the GKBA, even within the 2B approximation, is unable to go beyond the spectral features captured at the HF level. This is because the Landauer-B\"uttiker formula depends directly on the spectral function, Eq.~\eqref {eq:spfunc}, and thus on the retarded propagator. However, it is crucial to point out that the derivation of Eq.~\eqref {eq:LB} relies on the equality $G^<(\omega)=G^R(\omega)\varSigma^<(\omega)G^A(\omega)$. The latter holds in the steady state (in the absence of bound states) and it is a direct consequence of the Dyson equation Eq~\eqref{eq:propagator} with the full retarded self-energy. For this reason, it does not hold in the case of the GKBA-HF approximation.
This is a very important aspect for what follows. Eq.~\eqref{eq:curr-time} is derived directly from the definition of the particle current and contains explicitly the lesser Green's function, which in the GKBA-HF carries information about higher order correlation effect through a different  many-body self-energy compared to the one of the retarded propagator. Only if we employ the HF approximation for both the spectrum and the dynamics, then we would expect the GKBA to give the same result of the Dyson equation; consequently Eq.~\eqref{eq:curr} and Eq.~\eqref{eq:LB} would be equivalent. Instead, when higher order correlation effect (like in the case of the 2B self-energy) are included in the time evolution, we ought to rely on Eq.~\eqref{eq:curr-time} to capture features beyond the mean field approximation. In Ref.~\cite{Balzer2013gkba} similar reasoning was put forward for a broadened density-response spectrum within GKBA at the 2B level for finite systems.

\section {Results}
In the following, we will look at two different systems and compare the results of the stationary state of the GKBA-HF master equation with the solution of the stationary state of the full Dyson equation solved in the frequency domain. As we are not describing transient dynamics, we consider the partitioned approach~\cite{Stefanucci2004,Ridley2018JLTP} where the conducting device is suddenly brought in contact with the leads.

\subsection {Transport through a quantum wire}
In this section, we consider the case of two leads connected through a quantum wire. Electrons are assumed to be free in the leads whereas they experience a repulsive interaction inside the wire. To describe this system we work within the single-band Fermi-Hubbard model which is in turn tunnel-coupled with two infinite metallic leads. The total  adimensional Hamiltonian of the system reads:

\begin{eqnarray}
&\hat H_C =& \epsilon\sum_{i \sigma }^{}\hat n_{i \sigma }-\frac{1}{2}\sum_{\langle i,j\rangle\sigma}^{} \hat c^\dagger_{i \sigma } \hat c_{j \sigma }+U \sum_{i} \hat c^\dagger_{i \uparrow  } \hat c^\dagger_{i \downarrow} \hat c_{i \downarrow  } \hat c_{i \uparrow }\\
&\hat H_\alpha =& \sum_{k \sigma} \epsilon_{\alpha,k}  \hat d^\dagger_{\alpha, k \sigma} \hat d_{\alpha, k\sigma} \\ 
&\hat V_\alpha =& \sum_{i,k \sigma} T_{ik}^{\alpha} \hat c^\dagger_{i \sigma } \hat d_{\alpha,k \sigma}+ T_{ik}^{\alpha*} \hat d^\dagger_{{\alpha, k \sigma }} \hat c_{i \sigma },
\label {eq:hamFH}
\end{eqnarray}
where $\hat c^\dagger_{i\sigma},  \hat c_{i\sigma}$ are the creation (annihilation) operators of electrons in the basis labeled by $i$ and spin $\sigma=\uparrow/\downarrow$, with the sum running from $i=1$ to $i=N_x$, with $N_x$ denoting the length of the quantum wire. The operators $\hat d^\dagger_{\alpha, k \sigma},  \hat d_{\alpha,k \sigma}$ are the creation (annihilation) operators of the two different leads, denoted as $left$ and $right$, and labelled by $\alpha=L,R$. Hence, $\epsilon$ is the on-site potential, $U$ is the two-body interaction between spin-up and spin-down particle on the same site and $T_{ik}^{\alpha}$ is the tensor containing the coupling rates between the leads $\alpha$ and the chain.
In the following, we assume that the left and right leads are coupled to the first and last site of the chain respectively and described with the WBLA, i.e. $T^L_{ik}= T_L \delta_{i1}$ and $T^R_{ik}= T_R \delta_{iN_x}$. Thus, we have $\Gamma_{ij}=T_L^2 \delta_{i1}\delta_{ij} + T_R^2 \delta_{iN_x} \delta_{ij}$. Furthermore, the two leads are kept in a thermal state at the same temperature, i.e. $\beta_L=\beta_R$, but with different chemical potential given by $\mu_L=-\mu_R=\mu$. In what follows we have set $T_L^2=T_R^2=0.5$, $\mu=0.5$ and $U=1$.

\begin{figure}[t!]
	\includegraphics[width=0.9\linewidth]{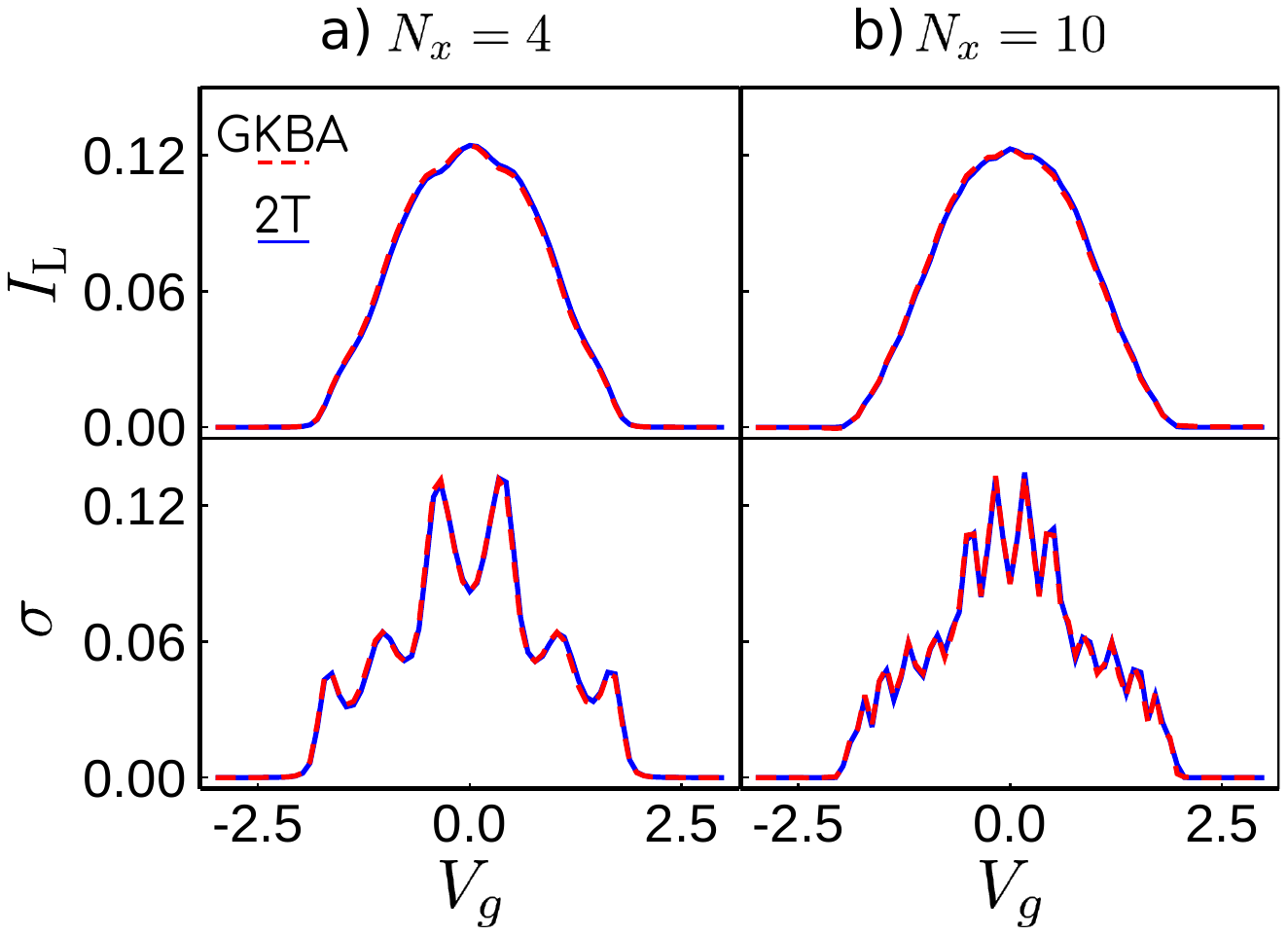}
	\caption{(Color online) Particle current in the left lead and conductance for a chain of a) $N_x=4$ and b) $N_x=10$ computed with the 2-times (solid blue) and the GKBA-HF master equation (red dashed) at the HF level.}
	\label{fig:1dhf}
\end{figure}
\begin{figure*}[t!]
	\includegraphics[width=0.9\linewidth]{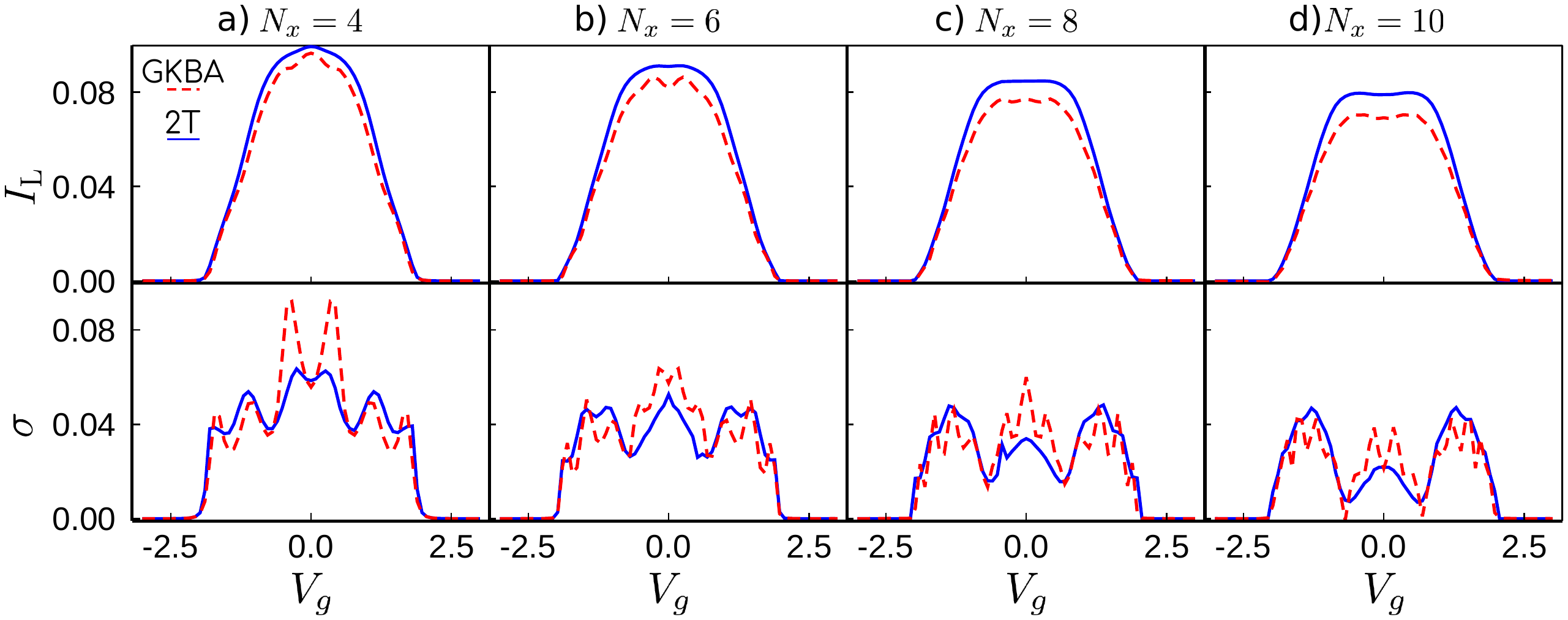}
	\caption{(Color online) Particle current in the left lead and conductance for a chain of a) $N_x=4$, b) $N_x=6$, c) $N_x=8$, d) $N_x=10$ computed with the 2-times (solid blue) at stationarity and the GKBA-HF master equation (red dashed) where the exchange self-energy is the Second Born one.}
	\label{fig:1d2b}
\end{figure*}

We use the gate potential $V_g = \epsilon + U/2$ to shift the spectrum of the central region with respect to the chemical potential of the leads and use the particle current through the wire act as a probe for the spectral properties of the central region at different energies.
Another physically relevant quantity is the differential conductance $\sigma=dI/dV$, where $V$ is the applied bias voltage across the central region,  i.e. $V=2\mu$. The conductance is more suited than the current, for the latter is an integrated quantity; instead the differential conductance is able to capture the details of the spectral weights.
In the closed system case, it is known that the full Dyson equation and the GKBA-HF give the same results. Nevertheless the  open system case is less trivial and it is worth to consider and show explicitly that the two coincide. The reason is purely technical and relies on the fact that we consider infinite leads and enforce the WBLA at the Hamiltonian level,contrary to other works which consider large but finite leads in the KBE approach~\cite{Latini2014} and therefore observe deviations between the two approaches even at the HF level.

In Fig.~\ref{fig:1dhf}, we show the currents (top panels) and the differential conductance (bottom panels) for two chains of $N_x=4$ and $N_x=10$ sites when the interaction is treated exclusively at the HF level.
The currents, as expected, show a perfect match, but it is only the differential conductance which reveals clearly the differences between the two systems and specifically highlights the spectral structures which are peculiar of the density of states of the two chains.
Moreover, we can appreciate the fact that both GKBA-HF and the Dyson equation approaches give the same results, thus confirming that even in the open-system case the GKBA-HF at the HF level returns exactly the same results as the 2-times one. Discrepancies would arise if structured couplings to the leads are used, namely if we drop the WBLA.

Moving forward to the second Born approximation for the self-energy in the collision integral, the solutions of the GKBA-HF master equation and the Dyson equation start to deviate from one another. In the upper panels of Fig.~\ref{fig:1d2b}, we display the stationary currents for different lengths of the quantum wire. With both methods, differences from the HF case emerge and two particular features are worth to be pointed out. The first one is that the current computed with the solution of the GKBA-HF master equation is always smaller than that of the full solution. This holds also as the size of the system is increased and actually the discrepancy increases. This might be due to the fact that the solution of the Dyson equations returns a more correlated state than the GKBA one due to the inclusion of the retarded component of the 2B self-energy in the Dyson equation for the retarded Green's function. The second one is that the GKBA-HF master equation shows sharper structures in the current which are more similar to the HF case in Fig.~\ref{fig:1dhf} than the one of the two-times one with the second Born self-energy.
Nevertheless, if we look at the conductance computed with both approaches (bottom panels in Fig.~\ref{fig:1d2b}) we notice that although the GKBA has indeed sharper peaks than the two-times one, it captures very well the main features of the differential conductance. Specifically,  we observe a reduction of the central structure as the size is increased and the rounding of the conductance peaks. Furthermore, the differential conductance is clearly different from the HF one.  It is important to point out that it has been shown elsewhere that the two-times solution computed at the second Born level introduced an excess in the damping of the oscillations in time for small system sizes~\cite{Hermanns2013,Hermanns2014}.  In frequency, this translates into an excess broadening in the peaks of the spectral function and therefore, for what we are concerned within this work, of the differential conductance. Indeed, as the system size also increases, the details of the structures become more similar.

Finally, from our simulations we conclude that in the one-dimensional case the solution of the GKBA-HF master equation is able to capture spectral features which go beyond the simple HF ones although the retarded Green's function contains only the HF propagator.

\subsection {Transport through a carbon nanotube}

The GKBA is computationally advantageous when compared to other NEGFs methods, specifically solvers of the two-time KBEs. Furthermore, the ansatz itself, as any other NEGFs approach, is virtually unaffected by the dimensionality of the system in exam. Therefore, it is worth to explore how the GKBA is able to capture spectral features in higher dimensional systems, where the phenomenology, given the topology of the coupling, is much richer than the one-dimensional case.

As a paradigmatic example of a two-dimensional system, we study the transport properties of a carbon nanotube (CNT), see Fig.~\ref{fig:CNT}. This choice is justified by the fact that graphene nanoribbons~\cite{Novoselov2005} and carbon nanotubes~\cite{Iijima1991} have been shown to be extremely sensitive to external perturbations~\cite{Gruner2006, Rodrigo2015, Rocha2015, Ruiz2016, Duffy2016, Ridley2017, Bian2018, Goldsmith2019, Tuovinen2019Nano}
making them optimal candidates for sensing technologies. In addition, it is known that disorder significantly influences
the operation of graphene-based
devices~\cite{Areshkin2007, Mucciolo2009, CastroNeto2009, Mucciolo2010, Dauber2014, Zhu2016, Ridley2019entropy}. When describing these
interesting nanoscale effects one must simultaneously take into account strong external fields, many-particle
interactions, and transient effects, for which the NEGF approach is suitable.

The total  adimensional Hamiltonian is the same as the one in Eq.~\eqref{eq:hamFH} with the exception of the Hamiltonian for the central system, which is now that of a Fermi-Hubbard model  defined on an honeycomb lattice:
\begin{equation}
\hat H_C = \epsilon\sum_{ {\bf i}\sigma}^{}\hat n_{ {\bf i}\sigma}-\frac{1}{2}\sum_{\langle {\bf i},{\bf j}\rangle_y \sigma}^{} \hat c^\dagger_{ {\bf i}\sigma} \hat c_{ {\bf j}\sigma}+U \sum_{i} \hat c^\dagger_{  {\bf i}\uparrow} \hat c^\dagger_{ {\bf i}\downarrow} \hat c_{  {\bf i}\downarrow} \hat c_{ {\bf i}\uparrow},
\label {eq:hamFHhoney}
\end{equation}
where now the indexes ${\bf i}$ run over the points of a honeycomb lattice and $\langle\cdots \rangle_y$ stands for the sum over the vertices of the lattice.
We consider periodic boundary conditions along the y direction as shown in Fig.~\ref{fig:CNT} and therefore a zigzag nanotube. In addition, we consider the number of armchair dimer lines in the transport direction $N_y=6$ representing a metallic character.
As in the one-dimensional example, the nanotube is coupled to two electronic leads within the WBLA, denoted as $left$  and $right$ and labelled as $L$ and $R$ respectively. Moreover, we assume that the left and right leads are diagonally coupled to the boundary sites of the carbon nanotube resulting in $\Gamma_{\mathbf {ij}}=T_L^2 \delta_{i_x1}\delta_{\mathbf {ij}} + T_R^2  \delta_{i_x N_x}\delta_{\mathbf {ij}}$. 
Furthermore, the two leads are kept in a thermal state at the same temperature, i.e. $\beta_L=\beta_R$, but with different chemical potential given by $\mu_L=-\mu_R=\mu$. Here we have used the same parameters as in the one-dimensional case, namely $T_L^2=T_R^2=0.5$, $\mu=0.5$ and $U=1$.

\begin{figure}[t!]
	\includegraphics[width=0.9\linewidth]{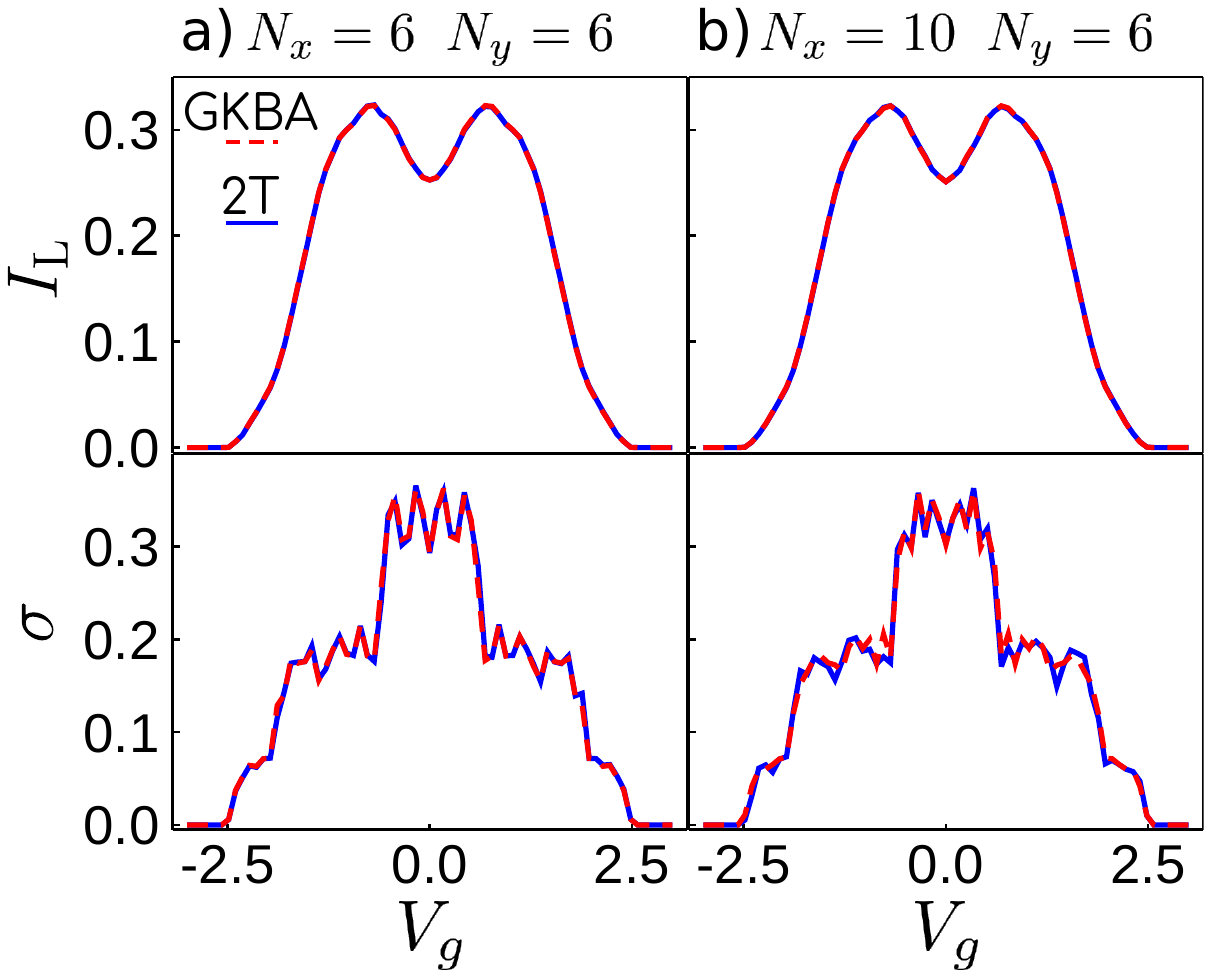}
	\caption{(Color online) Particle current in the left lead and conductance for a carbon nanotube with a) $N_x=6$, $N_y=6$, b) $N_x=10$, $N_y=6$ computed with the 2-times (solid blue) at stationarity and the GKBA-HF master equation (red dashed) with the exchange self-energy at the Hartee-Fock approximation.}
	\label{fig:2dhf}
\end{figure}

Also in this case, it is meaningful to first compare the results obtained with the GKBA master equation and the two-times solution at the HF level.  This comparison is shown in Fig.~\ref{fig:2dhf} for the case of a $N_x=6, \ N_y=6$ and $N_x=10, \ N_y=6$ CNTs. The agreement is perfect in both the current and the differential conductance, as it should be in this case since the solution of the GKBA-HF master equation corresponds to the exact solution of the Dyson equation.
Nevertheless, as previously done, this comparison serves to ensure that the presence of the leads in the GKBA-HF has been done appropriately and does not introduce any deviation between the two approaches.

\begin{figure}[t!]
	\includegraphics[width=0.9\linewidth]{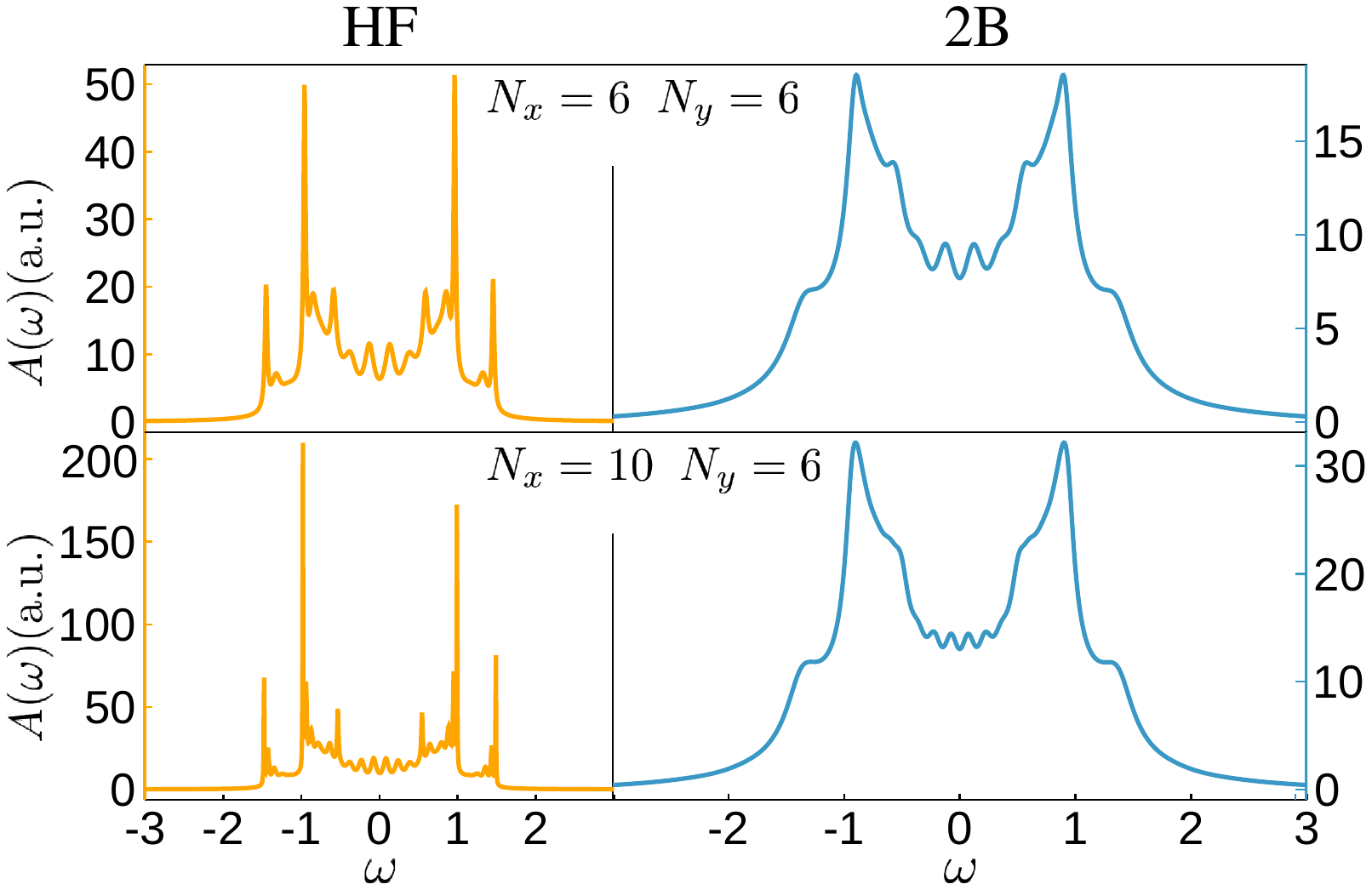}
	\caption{(Color online)  Density of states, shifted by $U/2$, of a carbon nanotube computed with the 2-times approach at the HF level (left) and  with the second-Born approximation (right). In the upper and lower panel,  we display the results for $N_x=6$, $N_y=6$ and $N_x=10$, $N_y=6$  respectively.}
	\label{fig:2ddos}
\end{figure}

The profile of the asymptotic currents shows the emergence of two peaks around the energies of $V_g \approx\pm 1$ with respect to the center of the band, signaling a concentration of states in this energy region. Indeed, the density of states $A(\omega)= \mathrm {Tr} \mathcal A(\omega)$, shown in Fig.~\ref{fig:2ddos} (left panel), confirms the presence of two structures at these energies, i.e. van Hove singularities~\cite{CastroNeto2009}. When the voltage of the gate $\epsilon$ is such that they enter the window of the leads, they produce an increase in the current flowing into or from the leads. 
In the bottom panel of Fig.~\ref{fig:2dhf}, we show the differential conductance and notice how it is larger instead when the window of the leads encloses the central part of the spectrum. This is due to the the fact that at low energies one has the maximum variation of the injected number of particles in the system. This conclusion comes from a closer inspection of the expression of the current in Eq.~\eqref{eq:curr} together with the definition of the differential conductance. In fact, the current contains two terms depending upon the bias, the lesser self-energy and the lesser Green's function of the central region. The change of the lesser self-energy on the bias is very weak and it is basically a shift of an otherwise constant function.
The lesser Green's function of the central region instead depends crucially on the bias because it carries information on the density of particles which changes drastically as the bias is changed. Therefore, at low energy, we have the most substantial variation of the number of particles because of a larger broadening when compared to higher energy states.

\begin{figure}[t!]
	\includegraphics[width=\linewidth]{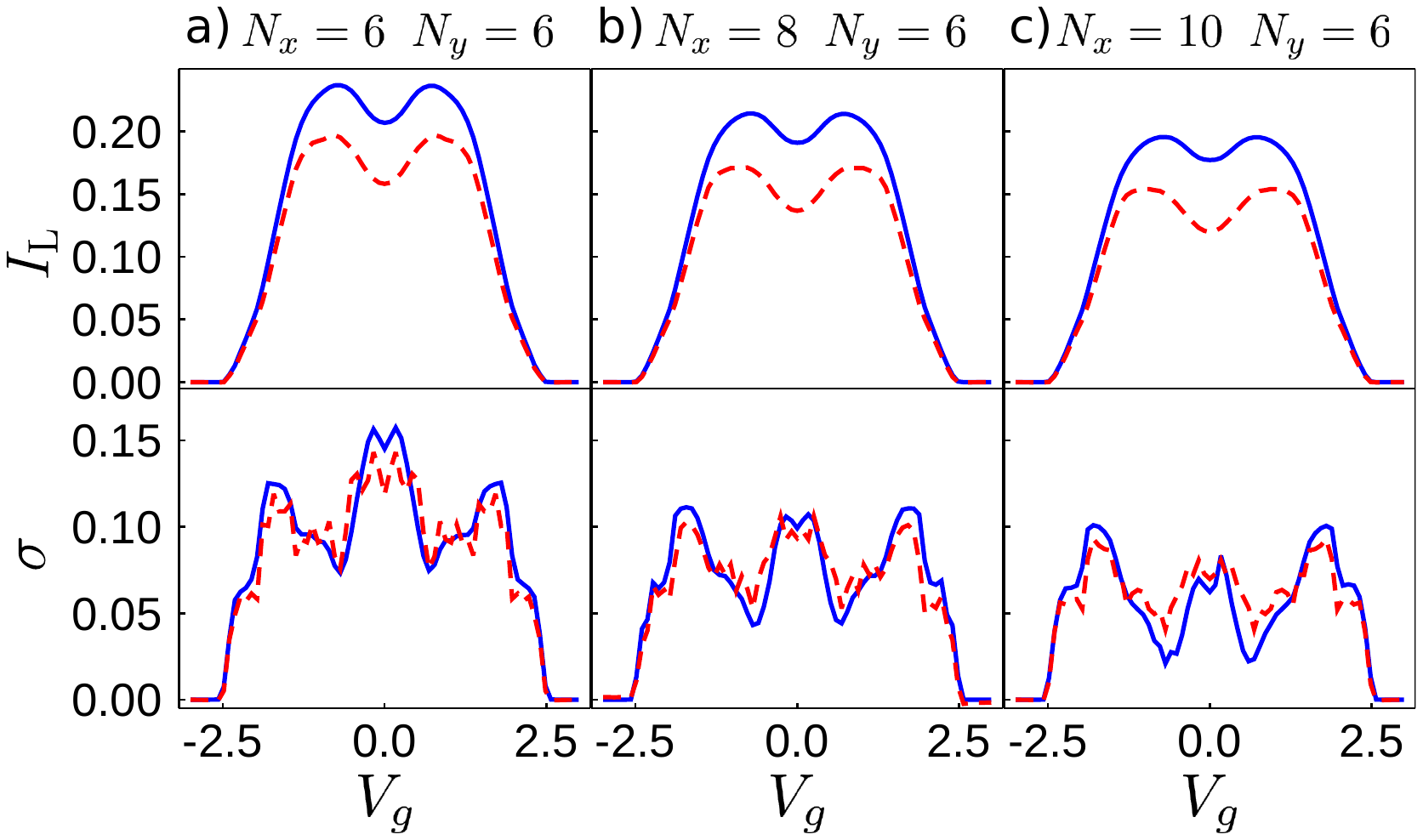}
	\caption{(Color online) Particle current in the left lead and conductance for a carbon nanotube with a) $N_x=6$, $N_y=6$, b) $N_x=8$, $N_y=6$, and c) $N_x=10$, $N_y=6$, computed with the 2-times (solid blue) at stationarity and the GKBA-HF master equation (red dashed) where the exchange self-energy is the Second Born one.}
	\label{fig:2d2b}
\end{figure}

In Fig.~\ref{fig:2d2b}, we show the asymptotic current and differential conductance calculated by including the 2B self-energy for different system sizes. As in the one dimensional case, the GKBA predicts a larger resistance with respect to the two-times, but both approaches agree on the fact that the current (top panels) reduces with the length of the nanotube. This is consistent with the expectation that by increasing the length of a non-interacting system  increases the scattering events and therefore reduces the current. In the HF case, the current remains the same even if the length of the system is increased and the transport remains fully ballistic.

The differential conductance, displayed in the bottom panels of Fig.~\ref{fig:2d2b}, shows instead a much better mutual agreement when compared to the one-dimensional case. In this case, the GKBA-HF is able to capture the two-times solution not only in a qualitative but also in a quantitative way. Specifically, it is able to capture the interaction-induced broadening which drastically changes the density of states of the system (shown in Fig.~\ref{fig:2ddos}) much better than in the one-dimensional case. There the GKBA appeared to retain the HF feature in a more pronounced way. 
Moreover, the differential conductance decreases with the length of the CNT, in agreement with the increase in the resistance discussed above, and experiences the emergence of a new interesting feature. As the length of the CNT is increased, the central peak of the conductance decreases faster than the two side structure. This suggests that the main role for the observed effect in the current is played by $G^<(\omega)$, which accounts for the change in the particle number in the system. In this case, a larger increase in the particle number at low energies results in a decrease of the differential conductance because of the repulsive interaction.

Finally, this behavior is a consequence of what we discussed in Sec.~\ref{sec:curr}, where we pointed out the difference between the two expressions for the current in Eq.~\ref{eq:curr} and Eq.~\ref{eq:LB}. The latter would lead to wrong results in the GKBA case, where the solution for the lesser Green's function is not related to the retarded one by the Dyson equation.
Instead, in Eq.~\ref{eq:curr}, the information on the correlations are carried by  the lesser Green's function through the presence of the 2B self-energy in the collision integral.

\section{Conclusions}
We have compared the solution of the Generalized Kadanoff-Baym Ansatz master equation with the stationary solution of the full two-times Dyson equation. Specifically, we studied a transport setup where two leads, considered in the wide band limit approximation, are connected through a central region, taken to be either a one-dimensional or a two-dimensional quantum system. In the first case, we considered a quantum wire and, in the second case, a carbon nanotube. In both setups, the fermionic particles experience a repulsive interaction treated with a second order Born approximation.

By using the stationary current and the conductance as figures of merit, we were able to show that the GKBA master equation, computed with a Hartree-Fock propagator and a second Born self-energy, is able to capture spectral features which the spectral function of the solution does not show. The latter is limited, by construction, to capture only features induced by the form of the propagator, and then by the Hartree-Fock self-energy.  

In other words, our findings give numerical evidence that the GKBA is able to capture spectral features beyond the HF propagator that are encoded in the lesser Green's function.  This aspect make of the GKBA a valuable tool for the simulation and the description of ARPES and pump-probe experiments, where the key object in the reconstruction of the signal corresponds exactly to the lesser component of the Green's function. In addition, even though we concentrated here on the stationary state, the present approach readily allows for studying also time-resolved transport in correlated quantum systems. This would be important in addressing transiently emerging phenomena, e.g., superconductivity and Majorana physics~\cite{Stefanucci2010, Jiang2011, Weston2015, Francica2016, Tuovinen2016PNGF, Thakurathi2017, Dehghani2017, Claassen2019, Tuovinen2019NJP}.

Our work, together with the recent speed-up achieved in the computation of the collision integrals for self-energies beyond the Second-Born one, contributes to show that the GKBA-HF is a powerful and reliable method to study out-of-equilibrium phenomena in many-body open and closed quantum systems.

\begin{acknowledgments}
F.C. acknowledges support from the ERC Synergy grant BioQ (Grant No. 319130), the EU projects HYPERDIAMOND (Grant No. 667192) and AsteriQs (Grant No. 820394), and the QuantERA project NanoSpin.
N.W.T. and N.L.G. acknowledge financial support from the Academy of Finland Center of Excellence program (Project no. 312058) and the Academy of Finland (Project no. 287750). R.T. acknowledges funding by the Academy of Finland Project No. 321540. N.L.G. acknowledges financial support from the Turku Collegium for Science and Medicine (TCSM).  The work has been performed under the Project HPC-EUROPA3 (INFRAIA-2016-1-730897), with the support of the EC Research Innovation
Action under the H2020 Programme; in particular, F.C. gratefully acknowledges the support of  Dr. Nicolino Lo Gullo, the department of Physics and Astronomy of the University of Turku, and the computer resources and technical support provided by the Finnish CSC. Numerical simulations were performed exploiting the Finnish CSC facilities under the Project no. 2000962 (``Thermoelectric effects in nanoscale devices'').
\end{acknowledgments}

\bibliography{gkbabib}

\end{document}